\documentclass[fdp,fleqn]{w-art}
\usepackage{times}
\usepackage{w-thm}
\usepackage[]{graphicx}

\usepackage{amssymb}

\begin{document}
\DOIsuffix{theDOIsuffix}
\Volume{55}
\Issue{1}
\Month{01}
\Year{2007}
\pagespan{1}{}



\title
{Consistent truncations with massive modes and holography}


\author[D. Cassani]{Davide Cassani\inst{1,}%
  \footnote{
  cassani at pd.infn.it$\;$ --$\;$ Speaker
                        }}
\address[\inst{1}]{Dipartimento di Fisica ``Galileo Galilei'',
		Universit\`a di Padova, Via Marzolo 8, 35131 Padova, Italy.}
\author[G. Dall'Agata]{Gianguido Dall'Agata\inst{1,2,}\footnote{dallagat at pd.infn.it}}
\address[\inst{2}]{INFN, Sezione di Padova, Via Marzolo 8, 35131 Padova, Italy.}
\author[A. F. Faedo]{Anton F. Faedo\inst{2,}\footnote{faedo at pd.infn.it}}
\begin{abstract}
We review the basic features of some recently found consistent Kaluza--Klein truncations including massive modes. We emphasize the general ideas underlying the reduction procedure, then we focus on type IIB supergravity on 5-dimensional manifolds admitting a Sasaki--Einstein structure,  which leads to half-maximal gauged supergravity in five dimensions. Finally, we comment on the holographic picture of consistency.
\end{abstract}
\maketitle                   






\section{Introduction}

In the context of the gauge/gravity duality, consistent Kaluza--Klein truncations of 10- and 11-dimensional supergravity have proved to be powerful solution-generating tools. A truncation of the higher-dimensional spectrum on some compact manifold is said to be consistent if all solutions of the truncated, lower-dimensional theory are also solutions to the full, higher-dimensional theory. Since they provide an effective lower-dimensional model with a restricted number of degrees of freedom, consistent truncations allow to tackle several problems in a setup which is much simpler than the original theory, guaranteeing at the same time the lifting of the solutions. As an example, consistent truncations to 5 dimensions provide a convenient framework for describing the renormalization group flow of the dual 4-dimensional quantum field theories, with the fifth coordinate playing the role of an energy scale.

A further, more formal, motivation to study consistent truncations is that they represent the main tool to investigate which lower-dimensional supergravities can be connected to string theory. 

Recently, there has been a revival of interest in consistent truncations, mainly motivated by the intense research activity towards a holographic description of strongly coupled condensed matter phenomena, such as superconductivity and quantum critical points exhibiting non-relativistic scale invariance.
A limitation of these holographic models is that most of them are {\it ad hoc} constructions built in a bottom-up approach, while in order to have full control on the gauge/gravity correspondence a rigorous embedding into string theory is needed. To achieve this, one should find a consistent truncation of 10-dimensional supergravity to the desired lower-dimensional model. A crucial, non-trivial feature required by these applications is that the truncation preserve massive and/or charged Kaluza--Klein modes (see in particular \cite{MaldMartTach, GauntlettKimVarelaWaldram, Gubser1, Gauntlett:2009dn}). 

Here, we will review the basic ideas underlying some recently found consistent truncations of higher-dimensional supergravity with massive modes. Often, the consistency of a truncation relies on some symmetry under which the preserved modes are invariant, and such that the truncated non-invariant modes are never generated in the equations of motion. In the cases of interest for us, the compact manifold admits a $G$-structure whose intrinsic torsion is also $G$-invariant. The modes preserved by the truncation are chosen to be precisely all the singlets under $G$. When, as it will be the case for us, among the invariant fields there is at least one spinor, and the invariant bosonic fields can be reconstructed by taking appropriate spinor bilinears, one can define a truncation ansatz which preserves a fraction of supersymmetry. A clear advantage in this case is that one disposes of a powerful organizing principle: since supersymmetric theories are very constrained, just a few data need to be provided in order to fully specify the truncated model.

In the following, we illustrate how these ideas work in practice by presenting a consistent truncation of type IIB supergravity on arbitrary squashed Sasaki--Einstein manifolds, leading to gauged $\mathcal N=4$ supergravity in five dimensions \cite{IIBonSE} (see also \cite{LiuEtAl, GauntlettVarelaSE, SkenderisTaylorTsimpis} for closely related work, and \cite{GauntlettKimVarelaWaldram, ExploitingN=2, BuchelLiu, GauntlettVarela07} for earlier supersymmetric results). As we will briefly describe, the resulting 5-dimensional model exhibits quite remarkable features, such as the presence of both massless and massive modes, as well as tensor fields dual to vectors charged under a non-abelian gauge group. Moreover, the retained Kaluza--Klein modes capture the universal pure gauge sector of the dual 4-dimensional super Yang--Mills theories.

\section{Type IIB supergravity on squashed Sasaki-Einstein manifolds}

A regular (respectively, quasi-regular) Sasaki--Einstein manifold $Y$ can be seen as a U(1) fibration over a K\"ahler--Einstein base manifold (respectively, orbifold) $B_{\rm KE}\,$:
\begin{equation}\label{SEmetric}
ds^2(Y) \,=\, ds^2(B_{\rm KE}) + \eta\otimes\eta\,,
\end{equation}
where $\eta$ is the globally defined 1-form specifying the fibration.
All 5-dimensional Sasaki--Einstein manifolds are also endowed with a real 2-form $J$ and a complex 2-form $\Omega$, both globally defined. These satisfy the algebraic constraints 
\begin{equation}\label{eq:AlgConstr}
\eta\,\lrcorner
\, J = \eta\,\lrcorner
\, \Omega = 0\,,\qquad \Omega\wedge J \,=\, \Omega\wedge \Omega\,=\,0\,,\qquad \Omega\wedge \overline\Omega \,=\,2\,J\wedge J \,=\,4\,{\rm vol}(B_{\rm KE})\,, 
\end{equation}
as well as the differential conditions
\begin{equation}
	\label{SEstructure}
d\eta \,=\, 2J\;,\quad\qquad 	d \Omega = 3 i\, \eta \wedge \Omega\,.
\end{equation}
The relations (\ref{eq:AlgConstr}) imply that the structure group of the 5-dimensional manifold, which generically is SO(5), is reduced to SU(2), with the forms $\eta$, $J$, $\Omega$ being SU(2) singlets. The conditions (\ref{SEstructure}) constrain the torsion of the SU(2) structure, which is required to also be an SU(2) singlet, and constant. Starting from the forms, the metric on the 4-dimensional subspace transverse to $\eta$ can be reconstructed by identifying a complex structure $I$ with respect to which $\Omega$ is of type $(2,0)$, and taking the product $J I$. We also have the Hodge duality relations
\begin{equation}\label{eq:*SEforms}
* \eta = {\rm vol}(B_{\rm KE})\;,\quad\qquad *J \,=\, J\wedge \eta\,,\qquad *\Omega \,=\, \Omega\wedge \eta\,.
\end{equation}

Our ansatz for the dimensional reduction is defined by writing down the most general expression for the metric and the various tensor fields of type IIB supergravity in terms of the forms characterizing the Sasaki--Einstein structure.
In doing this, we actually consider a class of internal metrics which is more general than (\ref{SEmetric}): we allow for an overall volume parameter (the ``breathing mode''), as well as for a parameter modifying the relative size of the U(1) fibre with respect to the size of the K\"ahler--Einstein base (the ``squashing mode''). 
Hence the class of spaces on which we are reducing is the one of {\it squashed} Sasaki--Einstein manifolds.

Specifically, our truncation ansatz for the 10-dimensional metric in the Einstein frame is \cite{MaldMartTach}:
\begin{equation}\label{eq:10dmetric}
ds^2 = e^{-\frac{2}{3}(4U+V)}ds^2(M)\,+\,e^{2U}ds^2(B_{\rm KE})\,+\,e^{2V}(\eta+ A)\otimes (\eta+ A)\,,
\end{equation}
where $A$ is a 1--form on the external 5-dimensional spacetime $M$, while $U$ and $V$ are scalars on $M$, which combined parameterize the breathing and the squashing modes of the compact manifold. 

Regarding the form fields of type IIB supergravity, let us for instance consider the NSNS 2-form $B$. We take the general expansion
\begin{equation}
	\label{eq:Bcov}
	B \,=\, b_2 + b_1 \wedge(\eta + A) + b^{J} J+ {\rm Re}(b^\Omega\,\Omega)\,,
\end{equation}
where $b_2$ is a 2-form, $b_1$ a 1-form, $b^J$ a real scalar, and $b^\Omega$ a complex scalar on $M$. The other type IIB form fields are expanded along the same lines. Finally, the dilaton $\phi$ and the Ramond-Ramond scalar $C_0$ are assumed to be independent of the internal coordinates.

The fact that the system of differential forms $\{\eta,\,J,\,\Omega\}$ is closed under the various operations appearing in the higher-dimensional equations of motion (exterior derivative, wedge product, Hodge star) ensures the consistency of the truncation. Indeed, one can plug the truncation ansatz in the 10-dimensional equations of motion, and check that they reduce to 5-dimensional equations; in particular, the dependence on the internal coordinates drops out. In parallel, one reduces the type IIB action by performing the integral over the internal space (care is needed in implementing the self-duality constraint on the Ramond-Ramond 5-form $F_5$, see discussion in \cite{IIBonSE}). Then one verifies that the resulting 5-dimensional action provides precisely the 5-dimensional equations that were obtained by reducing the 10-dimensional equations. This proves the consistency of the truncation.

As a final remark, we note that an ansatz expressed in terms of the invariant forms $\{\eta,\, J,\,\Omega\}$ of a generic SU(2) structure on the compact manifold would in general not be enough for having a consistent truncation. Indeed, while these forms are singlets under the structure group, their derivatives, providing the intrinsic torsion of the SU(2) structure, would generically contain non-singlet contributions, and this would a priori spoil consistency. Hence the simplicity of (\ref{SEstructure}), namely the fact that only singlets (with constant coefficients) appear on the right hand side, is as essential as (\ref{eq:AlgConstr}). Besides the case of Sasaki--Einstein structures in odd dimension, there are further examples of geometries characterized by a $G$-structure whose intrinsic torsion is also $G$-invariant, so that a truncation ansatz expressed in terms of the $G$-invariant tensors is consistent: for instance, the 7-dimensional weak-$G_2$ manifolds \cite{GauntlettKimVarelaWaldram}, the 6-dimensional Nearly-K\"ahler manifolds \cite{KashaniPoor:2007tr}, as well as the special holonomy manifolds, whose $G$-structure has vanishing torsion.

A situation in which the conditions just described are fulfilled is when the compact manifold is a coset space $\mathcal G/\mathcal H$. In this case, the structure group can be identified with $\mathcal H$, and the ansatz based on the singlets of the structure group can be rephrased as an ansatz invariant under the left-action of $\mathcal G$. Consistent truncations of supergravity on coset spaces have been studied in \cite{ExploitingN=2, T11reduction}. In \cite{T11reduction} (see also \cite{Bena:2010pr}), a specific example of 5-dimensional Sasaki--Einstein manifold was considered, namely the $T^{1,1}= (SU(2)\times SU(2))/U(1)$ coset space (also known as the base of the conifold). The consistent truncation outlined above can in this case be substantially enhanced, and provides the supersymmetric completion of a more limited non-supersymmetric truncation on $T^{1,1}$ \cite{PT-BHM}, containing several physically relevant conifold solutions, such as the Klebanov--Strassler and the Maldacena--Nu\~nez ones~\cite{KSandMN}.

\section{Gauged $\mathcal N=4$ supergravity description}

The 5-dimensional model stemming from the procedure described above can be understood in the framework of gauged $\mathcal N=4$ supergravity \cite{DHZ, SchonWeidner}. In this section we present just its most relevant features, referring to \cite{IIBonSE} for the expression of the complete bosonic action.\footnote{A detailed study of the inclusion of the fermionic sector has been performed in \cite{FermionicConsTrunc}.} Actually, thanks to the constraints dictated by half-maximal supersymmetry, in order to completely describe the model one needs to specify just a few data, namely the number of vector multiplets and the embedding tensor describing how the gauge group is embedded into the global symmetry group. Let us discuss them in turn.

The expectation of having $\mathcal N=4$ supersymmetry is first of all motivated by the gravitino ansatz. Type IIB supergravity contains two Majorana--Weyl gravitini of the same chirality $\Psi_M^{\alpha}$, where $\alpha=1,2$, and $M$ is a 10-dimensional spacetime index.
To define the truncation ansatz for these fields, we exploit the fact that the SU(2) structure condition implies the existence of two globally defined spinors $\zeta^{1}, \zeta^{2}$ on the internal manifold, being one the charge conjugate of the other. These are related to the forms $\eta,\,J,\,\Omega$ via appropriate spinor bilinears.
We use the two spinors to expand the 5-dimensional spacetime components of the 10-dimensional gravitini as
\begin{equation}
	\Psi^\alpha_\mu \,=\, \psi_\mu^{\alpha\,1} \otimes \zeta^1 + \psi_\mu^{\alpha\,2} \otimes \zeta^2\,.
\end{equation}
The 5-dimensional fields $\psi_\mu^{\alpha\,1},\,\psi_\mu^{\alpha\,2}$ can then be combined into four gravitini $\psi_\mu^i$, $i=1,\ldots,4$, satisfying the symplectic-Majorana condition
\begin{equation}
	\overline\psi{}_{\mu\,i} \equiv (\psi^i_{\mu})^{\dagger} \gamma^0= \Omega_{ij}(\psi_{\mu}^j)^TC\, ,
\end{equation}
where $\Omega_{ij}$ is the USp(4) invariant form while $C$ is the charge conjugation matrix. This provides the gravitino content of $\mathcal N=4$ supergravity.

Turning to the 5-dimensional bosonic sector, one can verify that it organizes in ${\cal N} = 4$ multiplets, with all the couplings respecting supersymmetry. For instance, by using a solvable parameterization one verifies that the target manifold of the scalar sigma-model is the prescribed homogeneous space
\begin{equation}
	{\cal M}_{\rm scal} = {\rm SO}(1,1) \times \frac{{\rm SO}(5,n)}{{\rm SO}(5)\times {\rm SO}(n)}\,,
\end{equation}
with $n=2$, which corresponds to a model with two vector multiplets. The counting and the couplings of the vector fields agree with this, though one has to take into account some complications due to the gauging. Indeed, while ungauged $\mathcal N=4$ supergravity in 5 dimensions contains eight vector fields, in our model we find four vectors and four 2-forms. The latter are seen as the Poincar\'e duals of the missing four vectors, the dualization being required by the gauging at hand.

In order to fully specify the gauging, one computes the embedding tensor mapping the gauge group generators into the generators of the duality group, which in the present case is SO(1,1) $\times$ SO(5,2). This determines the various additional couplings in the lagrangian with respect to the ungauged case, including the scalar potential, as well as the fermionic shifts appearing in the supersymmetry transformations (for the embedding tensor formalism we refer to \cite{Samtleben:2008pe} and references therein). In our case, the embedding tensor has components $f_{MNP}=f_{[MNP]}$ and $\xi_{MN}=\xi_{[MN]}$, where the indices $M,N,P=1,\ldots 7$ run in the fundamental of SO$(5,2)$. These can be determined by studying the gauge-covariant derivative of the scalars, and we find
\begin{equation}
\nonumber f_{125} = f_{256} = f_{567} = - f_{157} = -2,
\end{equation}
\begin{equation}\label{eq:OurEmbTensor}
\xi_{34}= -3\sqrt 2,\qquad\qquad \xi_{12}=\xi_{17}=-\xi_{26}= \xi_{67} = -\sqrt 2\, k\,.
\end{equation}
The higher-dimensional origin of the $f$-components is found in the geometric flux associated with the non-closure of the form $\eta$ on the internal manifold, while $\xi_{34}$ can be traced back to the non-closure of $\Omega$. 
The remaining non-zero $\xi$-components are proportional to the  constant $k$ parameterizing the internal Ramond-Ramond 5-form flux, $F_5^{\rm flux} = k\, J\wedge J\wedge \eta\,$.

By studying the commutation relations of the generators identified by the embedding tensor, one can infer that the gauge group is given by the product of U(1) with the three-dimensional Heisenberg group. The 10-dimensional origin of this gauge symmetry is found in part in the reparameterization invariance of the spacetime and in part in the shift symmetry of the type IIB form fields. 

Finally, we notice that taking the limit of vanishing fluxes ($d\eta = d\Omega = k = 0$), we obtain a consistent truncation of type IIB supergravity on $K3 \times S^1$ to ungauged $\mathcal N=4$ supergravity coupled to two vector multiplets.

\section{The holographic picture}

In this section we put our consistent truncation in the perspective of the gauge/gravity correspondence.

We start from the 5-dimensional scalar potential, which reads
\begin{equation}\label{eq:ScalPot}
\begin{array}{rcl}
\mathcal V &=&\displaystyle  -\,12 \,{ e}^{-\frac{14}{3}U-\frac{2}{3}V} +2\, { e}^{-\frac{20}{3}U + \frac{4}{3}V} + \frac{9}{2}\, { e}^{-\frac{20}{3}U-\frac{8}{3}V -\phi}|b^\Omega|^2 \\[4mm] 
 &&\displaystyle+\, \frac{9}{2}\, { e}^{-\frac{20}{3}U-\frac{8}{3}V+ \phi}|c^\Omega - C_0 b^\Omega|^2 + \;{ e}^{-\frac{32}{3}U-\frac{8}{3}V} \big[3\,{\rm Im}\big(b^\Omega\,\overline{c^\Omega}\big) + k\big]^2,
\end{array}
\end{equation}
where $c^\Omega$ is a complex scalar arising from the Ramond-Ramond 2-form potential. 
Studying $\mathcal V$ one finds two extrema. Choosing the RR flux $k=2$, these are
\begin{equation}
U \,=\, V \,=\, b^\Omega \,=\, c^\Omega \,=\, 0\,,\quad{\rm with}\;{\rm arbitrary}\; \phi, \;C_0\,,
\label{susyvacuum}
\end{equation}
and
\begin{equation}
e^{4U}=e^{-4V}=\frac{2}{3}\,,\qquad  b^\Omega=\frac{e^{i \theta+\phi/2}}{\sqrt{3}}\,\,,\qquad c^\Omega=b^\Omega  \tau\,, \qquad \tau \equiv (C_0  + i\,e^{-\phi})\,,
\label{nonsusyvacuum}
\end{equation}
where $\phi$, $C_0$ and $\theta$ are moduli of the solution. The cosmological constant $\Lambda \equiv \langle \mathcal V \rangle$ is negative in both cases ($\Lambda = -6$ for the first and $\Lambda = - \frac{27}{4}\,$ for the second), hence we have Anti-de Sitter vacua. The first extremum has $\mathcal N=2$ supersymmetry, and lifts to the standard AdS$_5\times\,$Sasaki--Einstein$_5$ solution of type IIB supergravity.
The second extremum is non-supersymmetric, and corresponds to an AdS$_5$ solution of type IIB supergravity with squashed internal metric, originally found in \cite{RomansIIBsols}. 

By studying the mass spectrum of the field fluctuations about these backgrounds, one can implement the standard AdS/CFT dictionary and deduce the anomalous dimensions of the dual CFT operators. For the non-supersymmetric extremum we find some irrational conformal dimensions;  since the square masses are all positive, the vacuum is stable at least with respect to the modes kept in the truncation. The results for the supersymmetric vacuum are summarized in table \ref{GaugeGravityTable}. By studying the dual spectrum (for instance in the well-known cases of $S^5$ and $T^{1,1}$), we see that we are keeping just flavor singlets, which are built in terms of the gauge superfield $W_\alpha$ in the $\mathcal N=1$ super Yang-Mills theory. Since this is a fermionic superfield, we can construct just a finite number of non-vanishing combinations, which precisely match the degrees of freedom of the gravity model. We conclude that our truncation describes the large $N$ limit of the universal gauge sector of $\mathcal N=1$ super Yang--Mills theories in 4 dimensions. 

As noticed in \cite{SkenderisTaylorTsimpis}, on the field theory side the consistency of our truncation translates into the fact that the set of operators appearing in table~\ref{GaugeGravityTable} is closed under the operator product expansion (at least in the large $N$ limit). It would be interesting to explore to what extent this is a general feature of consistent truncations with a field theory dual.

\renewcommand{\arraystretch}{1.2}
\begin{table}
\begin{center}
$	\begin{array}{rcccccccl}\hline
\mathcal N=2\: {\rm multiplet} && {\rm field\: fluctuations} && m^2 &\phantom{s}& \Delta & \phantom{s}& {\rm dual\; operators}  \\\hline
\rule{0pt}{3ex}
{\rm gravity} && \begin{array}{c} A - 2a_1^J \\ g_{\mu\nu} \end{array}&& \begin{array}{c} 0 \\ 0 \end{array} && \begin{array}{c} 3\\ 4 \end{array} &&  {\rm Tr}(W_{\alpha}\overline W_{ \dot\alpha})+\ldots\\ \hline
{\rm universal \;hyper} && \begin{array}{c} b^\Omega - i \,c^\Omega \\ \phi\,,\;\;C_0 \end{array} && \begin{array}{c} -3 \\ 0 \end{array} && \begin{array}{c} 3 \\ 4 \end{array} && {\rm Tr} (W^2) +\ldots   \\ \hline
		&& b_1,\;c_1 && 8 && 5 && \\
{\rm massive\; gravitino} && a_2^\Omega && 9 && 5 &&  {\rm Tr}(W^2\overline{W}_{\dot \alpha}) + \ldots \\ 
&&		b_2,\;c_2 && 16 && 6  && \\ \hline
{\rm massive \;vector} && \begin{array}{c} U - V\\ A + a_1^J\\ b^\Omega + i \,c^\Omega \\ 4U + V \end{array}&&\begin{array}{c} 12\\ 24 \\ 21\\ 32  \end{array}  &&\begin{array}{c}  6\\   7 \\  7\\  8 \end{array} &&  {\rm Tr}(W^2\overline W{}^2) + \ldots \\ \hline
\end{array}$
\caption{Mass eigenstates of the type IIB fields in our truncation on the supersymmetric AdS$_5\times\,$Sasaki--Einstein$_5$ background, and their dual superfield operators. Since the vacuum has $\mathcal N=2$ supersymmetry, the field fluctuations organize in $\mathcal N=2$ multiplets. We provide the mass eigenstates entering in each multiplet, together with their mass eigenvalues $m^2$, the conformal dimension $\Delta$ of the corresponding dual operators, and the dual superfields accommodating the single operators. The mass eigenvalues are evaluated choosing the flux $k=2$, which yields a unit AdS radius. 
The vectors $b_1,\,c_1,\, A + a_1^J$ acquire a mass via a St\"uckelberg mechanism (the notation is the one of \cite{IIBonSE}).} \label{GaugeGravityTable}
\end{center}
\end{table}

\begin{acknowledgement}
The authors are supported in part by the ERC Advanced Grant no. 226455, ``Supersymmetry, Quantum Gravity and Gauge Fields'' (\textit{SUPERFIELDS}) and by the Fondazione Cariparo Excellence Grant ``String-derived supergravities with branes and fluxes and their phenomenological implications''.\end{acknowledgement}


\end{document}